\documentclass[sn-mathphys-num]{sn-jnl}

\usepackage{graphicx}%
\usepackage{multirow}%
\usepackage{amsmath,amssymb,amsfonts}%
\usepackage{amsthm}%
\usepackage{mathrsfs}%
\usepackage[title]{appendix}%
\usepackage{xcolor}%
\usepackage{textcomp}%
\usepackage{manyfoot}%
\usepackage{booktabs}%
\usepackage{algorithm}%
\usepackage{algorithmicx}%
\usepackage{algpseudocode}%
\usepackage{listings}%
\usepackage{makecell}
\usepackage{soul} 
\usepackage{ulem}
\usepackage{amssymb}

\begin{document}

\title[Article Title]{Personalized News Recommendation with Multi-granularity Candidate-aware User Modeling}


\author{\fnm{Qiang} \sur{Li}}

\author[1]{\fnm{Xinze} \sur{Lin}}

\author[1]{\fnm{Shenghao} \sur{Lv}}

\author*[2]{\fnm{Faliang} \sur{Huang}}
\email{hfl@nnnu.edu.cn}

\author*[1]{\fnm{Xiangju} \sur{Li}}
\email{xiangjul@sdust.edu.cn}


\affil[1]{\orgdiv{Shandong University of Science and Technology}, \orgname{College of Computer Science and Engineering}, \orgaddress{\city{Qiangdao}, \postcode{266500}, \country{China}}}

\affil[2]{\orgdiv{Nanning Normal University}, \orgname{Guangxi Key Lab of Human-machine Interaction and Intelligent Decision}, \orgaddress{\city{Nanning}, \postcode{530100}, \country{China}}}


\abstract{Matching candidate news with user interests is crucial for personalized news recommendations. Most existing methods can represent a user's reading interests through a single profile based on clicked news, which may not fully capture the diversity of user interests. Although some approaches incorporate candidate news or topic information, they remain insufficient because they neglect the multi-granularity relatedness between candidate news and user interests. To address this, this study proposed a multi-granularity candidate-aware user modeling framework that integrated user interest features across various levels of granularity. It consisted of two main components: candidate news encoding and user modeling. A news textual information extractor and a knowledge-enhanced entity information extractor can capture candidate news features, and word-level, entity-level, and news-level candidate-aware mechanisms can provide a comprehensive representation of user interests. Extensive experiments on a real-world dataset demonstrated that the proposed model could significantly outperform baseline models\footnote{The source code of this work will be released at https://github.com/pznger/MGCA-Codes.}.}

\keywords{news recommendation, multi-granularity candidate-aware, word-level, entity-level}



\maketitle

\section{Introduction}\label{sec1}

With the rapid expansion of the Internet, users have increasingly relied on platforms such as Microsoft News and Google News for news consumption. However, the sheer volume of news generated daily makes it challenging for users to efficiently obtain the content of interest. Personalized news recommendations address this issue by directing users to relevant content, thereby mitigating the information overload. Consequently, personalized recommendations play a crucial role in online news platforms and have attracted significant attention from both the industry and academia~\cite{DBLP:conf/acl/AnWWZLX19,QiWWYY0020,DBLPZhuCLYLQZ22,TranSZTK23,WangWLPZZQ23,XuPLSW24,DBLPHuangHXL22}.

\begin{figure*}[!t]
\centering
\includegraphics[width=1.02 \columnwidth]{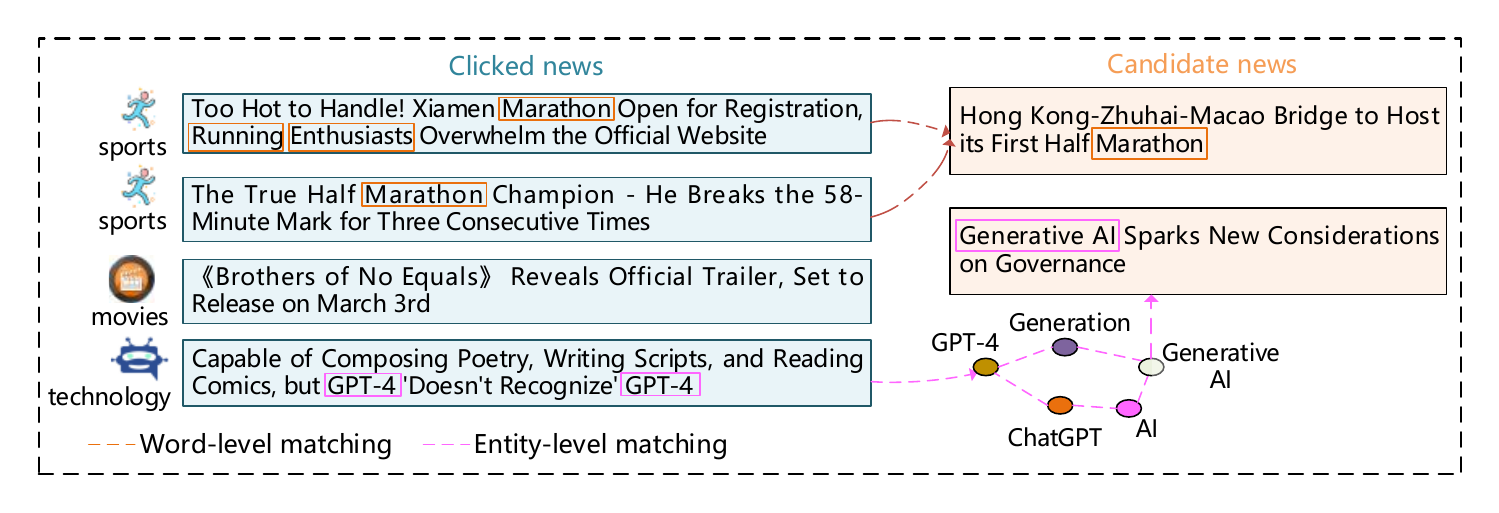} 
\caption{An example of user reading behaviours with four clicked news and two candidate news. 
}
\label{fig1}
\end{figure*}

The main challenge in personalized news recommendations is to accurately match candidate news with user interests. Most existing methods rely on deep learning techniques to independently model candidate news based on textual information and infer user interests from their click history~\citep{OkuraTOT17,WuWGQHX19}. 
For example, \citet{OkuraTOT17} used an autoencoder to extract semantic features from news and a GRU network to capture user interests based on historical click sequences. NPA~\citep{DBLP:conf/kdd/WuWAHHX19} independently learns news representations through a word-level personalized attention network and user-interest representations through a user-level personalized attention network, matching them via the inner product of user interests and candidate news representations. However, these methods can model user interests in a candidate-agnostic manner, overlooking the impact of candidate news on user modeling. In reality, users typically have multiple interests, and without considering candidate news during user modeling, it becomes challenging to accurately match news to specific interests~\citep{DKNWangZXG18}. For instance, the first candidate news item in Figure~\ref{fig1} can be related to marathons, whereas users have diverse interests in sports, technology, and movies. This candidate news only aligns with the sports interest of users, and they may be primarily interested in the keyword ``\textit{Marathon}''. Consequently, independently modeling user interests and candidate news can result in suboptimal and less effective matching.

Some researchers have enhanced the alignment between candidate news and user interests by integrating candidate news into user modeling. For instance, \citet{DKNWangZXG18} developed a candidate-aware attention mechanism that learns user interest representations by aggregating historically clicked news based on its relevance to candidate news. Similarly, \citet{QiWWH22} introduced a candidate-aware CNN network to incorporate candidate news into local behavior context modeling, facilitating the learning of candidate-aware short-term user interests. Although these models are promising, most only integrate candidate news into user-interest modeling at the news level.
The interactions between clicked news and candidate news at the word and entity levels offer detailed clues for a better understanding of user interests. For instance, as shown in Figure~\ref{fig1}, the first clicked news includes terms such as ``\textit{Marathon}'',``\textit{Running}", and ``\textit{Enthusiasts}'', which relate to the word ``\textit{Marathon}'' in the first candidate news. Similarly, the second clicked news shares word-level relatedness with the candidate news through the term ``\textit{Marathon}''. This similarity suggests that the user may be interested in candidate news. Additionally, leveraging knowledge graphs for entity matching between clicked and candidate news provides deeper insight into user interests. For example, entity ``\textit{GPT-4}'' in the third clicked news is related to the ``\textit{Generative AI}'' in the second candidate news. GPT-4 is a component of ChatGPT, which is a representative AI technology. Furthermore, GPT-4 also is a generation method. Based on entity-level matching, the user is interested in the second candidate news. Therefore, utilizing the relatedness between clicked news and candidate news at both word and entity levels is advantageous for interest alignment.

Based on our observations, a \underline{m}ulti-\underline{g}ranularity \underline{c}andidate-\underline{a}ware user modeling framework (MGCA) was proposed for personalized news recommendations. This framework integrated word-, entity-, and news-level granularity information from candidate news into user modeling for more precise matching. At the word level, we developed a candidate-aware Fastform network that used candidate news representation as a query to guide the modeling of word-level relatedness between clicked news and candidate news. At the entity level, an attention mechanism captured the entity-level relationship between candidate and clicked news. At the news level, a candidate-aware transformer network learned the coarse granularity matching features. The fusion of these three levels of granularity enhanced the representation of user interests. The main contributions of this study are summarized as follows.

\begin{itemize}
    \item  The multi-granularity correlations between user interests and candidate news were investigated, and relevant matching features were extracted.
    \item An MGCA framework was proposed to learn candidate-aware user interest features across multiple granularities, such as word, entity, and news levels, through mechanisms specifically designed to align user interests with candidate news.
    \item The experimental results demonstrated that MGCA outperformed previous personalized news recommendation methods in terms of AUC, MRR, nGG@5, and nGG@10.
\end{itemize}

The remainder of this paper is organized as follows. Section~\ref{Related work} presents a systematic review of existing studies on personalized news recommendations. Section~\ref{method} formally defines the personalized news recommendation task and details the proposed MGCA model. Section~\ref{exper} provides information about the adopted dataset, followed by a discussion of the comprehensive evaluation, ablation studies, and a case analysis of our model. Finally, Section~\ref{conc} presents the conclusions.

\section{Related work} \label{Related work}
Personalized news recommendations were applied to identify the engaging news for each user and enhance the reading experience. This could be crucial for online news services and has been extensively explored in recent years~\citep{WangWLX20,DBLP:conf/acl/AnWWZLX19,DKNWangZXG18,DBLP:conf/recsys/LiuLWQCS020}. 

Compared with other recommendation domains~\citep{LiangLSZLZ24,WangCGFL24,DBLPChenLLLD24}, news recommendations can involve greater complexity in text mining. Conventional content-based approaches~\citep{contentRecom,DBLP:conf/www/LiCLS10} can utilize news term frequency features, such as TF-IDF, alongside the user profiles derived from historically clicked news to recommend news with a higher similarity to the user’s profile. Content-free collaborative filtering methods~\citep{DBLP:conf/www/DasDGR07,SongWFZQY14} have also been applied to predict ratings based on past ratings from current or similar users~\citep{conf/icml/MarlinZ04}. However, both approaches have limitations. Content-based methods struggle to effectively model news and user profiles, whereas collaborative filtering methods often face the cold-start problem owing to frequent news substitution. Hence, hybrid methods~\citep{LiWLKP11,MoralesGL12} that combine the strengths of both approaches have been proposed to improve the news recommendation performance.

In recent years, significant progress has been made by deep learning neural networks such as GAT~\citep{GAT}, Transformers~\citep{Transf}, and Fastformers~\citep{Fastf} in learning semantic features from text, leading to notable success in various NLP tasks~\citep{HuangLYZZ,HuangYBLLW22} and recommendation systems~\citep{WangGRMZ23,BansalGK24}. Therefore, deep learning techniques have been employed in some studies to generate semantic representations of news with high effectiveness in capturing intricate patterns and meaningful information from textual content. The effective alignment of user interests with candidate news has been recognized as a crucial component for successful recommendations, resulting in the development of methods that focus on modeling both user interests and candidate news to enhance the recommendation process. For instance, embeddings are learned from news bodies using an auto-encoder, and user interests are modeled based on clicked news using a Gated Recurrent Unit (GRU) network, as demonstrated by~\citet{OkuraTOT17}. Personalized matching scores for news ranking can be computed using the dot product of the user and news embeddings. Similarly, \citet{WuWGQHX19} utilized multi-head self-attention networks to generate embeddings for news content from titles and model user interests from clicked news. \citet{FUM} emphasized the importance of word-level interactions among various clicked news items from the same user, as these interactions provided crucial insights into user interests. To address this, they proposed a fine-grained and efficient user modeling framework that could capture user interests through detailed behavioral interactions for improved news recommendations. Despite their commendable performance, these methods often ignore the inadequacy of separately modeling candidate news and user interests to obtain a precise alignment between them.

For more precise interest matching, integrating candidate news information into the user interest modeling process is essential~\cite{QiWW021,QiWWH22,WangWLX20}. \citet{WangWLX20} suggested enhancing user representation by performing fine-grained matching between candidate news and clicked news of users, leading to more accurate and detailed modeling of user interests and improved news recommendation performance. \citet{QiWW021} introduced a knowledge-aware interactive matching technique that employed an entity co-attention network and a semantic co-attention network to model the interactions between candidate news and clicked news. Additionally, \citet{QiWWH22} used candidate news as a reference to model candidate-aware global user interests and incorporated it into local behavior context modeling, capturing candidate-aware short-term user interests. This approach refined the recommendation process by addressing both long-term preferences and recent interactions with candidate news.

The analysis of existing approaches revealed that some methods independently model user interests while neglecting the effects of candidate news. Although other methods can incorporate candidate news into user interest modeling, they often focus only on fine-grained or coarse-grained interactions between candidate news and clicked news, overlooking the benefits of multi-granularity matching features. In contrast, we introduced a multi-granularity candidate-aware user modeling framework that efficiently leveraged word-, entity-, and news-level granularity mechanisms. This framework captured the matching features at different levels between user-clicked news and candidate news, ultimately leading to more informative news recommendations.

\section{Methodology}\label{method}

\begin{figure*}[!t]
\centering
\includegraphics[width=1.0 \columnwidth]{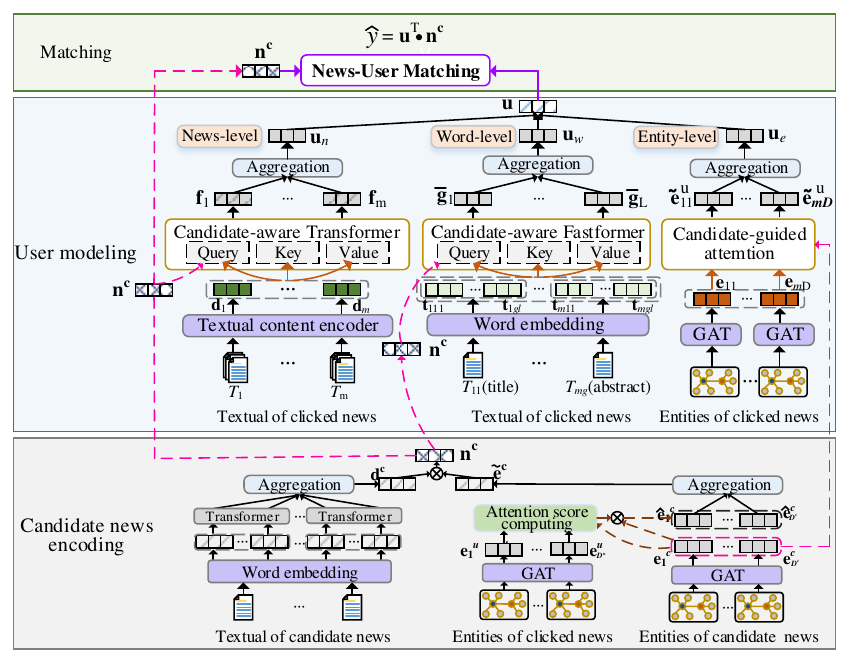} 
\caption{The framework of MGCA.
}
\label{frame}
\end{figure*}

This section first defines the task of personalized news recommendations and then provides a detailed description of the proposed MGCA model for this task, with the overall framework illustrated in Figure \ref{frame}. The key notations used in this study are listed in Table \ref{notation}.

\begin{table}[!t]
\centering
\caption{Notations and their meanings.}
\label{notation}
\setlength{\tabcolsep}{1mm}{
\begin{tabular}{p{2.5cm}p{10cm}}
\hline
Symbol & Description  \\
\hline
$n^c$ & the candidate news. \\
$c_i$ & the \textit{i}-th user clicked news.  \\
$m$, $g$, $l$ & the number of user clicked news, the number of genres of textual information and the number of tokens of each genre textual.\\
$T_i$, $T_i^c$ & the 
\textit{i}-th genre of textual information in the clicked news \textit{c} and the \textit{i}-th genre of textual information in the candidate news $n^c$, respectively.   \\
$t_{ij}$, $t_{ij}^c$ & the \textit{j}-th token in the sequence $T_i$ and $T_i^c$, respectively.  \\
$D^c$, $D$, $D^u$ & the number of entities of the candidate news, each clicked news, and all the clicked news.  \\
$\mathbf{d}^c$, $\tilde{\mathbf{e}}^c$, $\mathbf{n}^c$ & the textual representation, entity representation and the final representation of the candidate news, respectively.  \\
$\mathbf{u} _w$, $\mathbf{u}_e$, $\mathbf{u}_n$, $\mathbf{u}$ & the word-level, entity-level, news-level and the final user interest representation, respectively. \\
\hline
\end{tabular}}
\end{table}

\subsection{Task Definition}

Given a user $u$ and a candidate news item $n^c$, the task is to calculate a relevance score $\hat{y}$ that quantifies the user’s interests in the content of $n^c$. Subsequently, the candidate news are ranked and recommended to the user based on their respective relevance scores.

User $u$ is associated with $\{c_1, c_2, \dots, c_m\}$ that represents the set of previously clicked news. Let $E = \{e_1, e_2, \dots, e_{D^u}\}$ represent the set of entities in news $c$, and $e_i$ be the \textit{i}-th entity of $c$.
A news item $c$ can be considered to typically contain g genres of textual information, such as titles and abstracts, which is denoted as $T = \{T_1, T_2,\dots, T_g\}$, where $T_i$ corresponds to the i-th category of text. $T_i = \{t_{i1}, t_{i2}, \dots, t_{il}\}$ denotes the textual sequence consisting of multiple words, where $t_{ij}$ is the $j$-th word in the sequence $T_i$; and $l$ represents the maximum sequence length.

\subsection{Model Description}

In this section, the proposed MGCA model for news recommendations is described in detail. The MGCA model comprised three main modules: (1) Candidate News Encoding, which derives the representation of candidate news using both a textual information extractor and an entity information extractor; (2) User Modeling, which learns user interest representation through a multi-granularity candidate-aware mechanism that captures word-level, entity-level, and news-level interactions between candidate news and clicked news; and (3) Candidate News with User Interest Matching, which computes the probability that the target user will click on the candidate news by leveraging the matching representations between the news and user interests. 

\subsubsection{Candidate News Encoding}\label{candid}
Candidate news encoding was designed to extract features from candidate news through two main steps: (1) the textual information extractor that learned the news embeddings from the text of the news and (2) the entity information extractor that incorporated the knowledge graph information into the entity modeling to obtain rich information about the entities. The resulting information from these two steps was then fused to form the final candidate news representation, which was utilized for multi-granularity candidate-aware user modeling and click prediction. 

\textbf{Textual Information Extractor.} 
To differentiate between candidate news and user-clicked news, the superscripts “c” and “u” were used to denote the representations associated with the candidate news and user-clicked news, respectively. 
Each word $t^c_{ij}$ in the $i$-th genre of the candidate news text $T_i^c$ was first mapped into a $d_w$-dimensional vector $\mathbf{t}_{ij}^c \in \mathbb{R}^{d_w}$, where $d_w$ is the embedding size dimension. Consequently, text $T_i^c$ can be represented as a feature map $\mathbf{T}_i^c = [\mathbf{t}_{i1}^c; \mathbf{t}_{i2}^c; \dots; \mathbf{t}_{il}^c] \in \mathbb{R}^{l \times d_w}$, $[\cdot ; \cdot]$, which involves concatenating vectors along columns. 

Transformer architecture, based on self-attention mechanisms, has demonstrated superior performance in capturing intricate semantic and syntactic relationships among words. Therefore, the transformer was used to model different genres of textual information. The contextualized word representations of the \textit{i}-th genre of textual information $\hat{\mathbf{T}}_i^c$ are represented as follows:
\begin{equation}\label{eta}
\begin{split}
\hat{\mathbf{t}} _{i1}^{c}, \hat{\mathbf{t}}_{i2}^{c}, \ldots, \hat{\mathbf{t}}_{il}^{c} =Trans (\mathbf{t}^c_{i1},\mathbf{t}^c_{i2}, \ldots, \mathbf{t}^c_{il})\\
\hat{\mathbf{T}}_i^c = [\hat{\mathbf{t}}_{i1}^{c}; \hat{\mathbf{t}}_{i2}^{c}; \ldots; \hat{\mathbf{t}}_{il}^{c}]
\end{split}
\end{equation}\label{eta}
where $\hat{\mathbf{T}}^{c}_i \in \mathbb{R}^{l \times d_w}$, 
$Trans(\cdot)$ represents the architecture of Transformer.
Motivated by \cite{DBLP:conf/www/GeWWQH20},
genre-specific news embeddings were attentively aggregated to derive the overall news embedding. The $i$-th genre text representation $\hat{\mathbf{t}^c_i} \in \mathbb{R}^{d_w}$ can be generated as follows: 
\begin{equation}
    \hat{\mathbf{t}^c_i} = softmax( \mathbf{W}_1 \cdot \mathbf{(\hat{\mathbf{T}}}^c_i)^{T} + \mathbf{b}_1) \cdot \mathbf{\hat{\mathbf{T}}}^c_i
\end{equation}
where $\mathbf{W}_1 \in \mathbb{R}^{d_w}$ and $\mathbf{b}_1 \in \mathbb{R}^{l}$ denotes the trainable parameters; $( \cdot )^{T}$ represents the transpose operation; $softmax(\cdot)$ represents the softmax function.

The representations of all genres of the news text were concatenated to form a textual representation of the candidate news:
\begin{equation}\label{eta}
\mathbf{d}^c= MLP(\hat{\mathbf{t}}_1^c \oplus \hat{\mathbf{t}}_2^c \oplus \ldots \oplus \hat{\mathbf{t}}_g^c)
\end{equation}\label{eta}
where $\mathbf{d}^c \in \mathbb{R}^{d_w}$, $\oplus$ denotes the the operation of concatenating vectors.

\textbf{Entity Information Extractor}.
News often includes entities that convey key messages and facilitate the direct comprehension of content. Knowledge graphs storing the extensive external information about entities could offer a promising research direction for enhancing the understanding of news. For each entity $e^c_i$ in the candidate news, the initial entity embedding vectors $\mathbf{e}^c_i$ can be obtained using TransE, which can be pretrained on the knowledge graph. Recognizing that an entity is represented not only by its own embeddings but also by those of its neighboring entities, a graph attention network is employed to capture entity representations~\citep{QiWW021}. Consequently, the high-level entity representation $\mathbf{e}_i^c \in \mathbb{R}^{d_e}$ of entity $e_i^c$ in candidate news was obtained. Similarly, the entity representation $\mathbf{e}_i^u \in \mathbb{R}^{d_e}$ of entity $e_i^u$ in the user’s clicked news was derived.

The entities within candidate news can often vary in informativeness for relevance matching. To capture the critical entity information, a clicked-news-guided attention mechanism was employed. Specifically, the relevance score between entities $e_i^c$ and $e_j^u$ was first calculated using the following measurement function:
\begin{equation}
    \alpha_{ij} = \frac{\mathbf{e}_i^c \cdot \mathbf{e}_j^u}{\sum_{i=1}^{D^c} \mathbf{e}_i^c \cdot \mathbf{e}_j^u}
\end{equation}
where $i \in \{1,2,\dots, D^c\}$, $j\in \{1,2,\dots, D^u\}$; $D^c$ denotes the number of entities of the candidate news; and $D^u$ is the number of entities of the user's clicked news. Then, the clicked news weighted entity representation $\hat{\mathbf{e}}_i^c \in \mathbb{R}^{d_e}$ of entity $e_i^c$ can be obtained as follows:
\begin{equation}
   \hat{\mathbf{e}}_i^c = \sum\limits_{j=1}^{D^u} \alpha_{ij} \mathbf{e}_i^c
\end{equation}
The entity information $\tilde{\mathbf{e}}^c \in \mathbb{R}^{d_e}$ of the candidate news can be obtained by concatenating all entities:
\begin{equation}
   \tilde{\mathbf{e}}^c = MLP( \hat{\mathbf{e}}_1^c \oplus \hat{\mathbf{e}}_2^c \oplus \dots \oplus \hat{\mathbf{e}}_{D^c}^c)
\end{equation}

Textual representation and knowledge-enhanced entity representation are two crucial types of information for candidate news. Therefore, we aggregated these representations to form an overall candidate news representation:
\begin{equation}
    \mathbf{n^c}=\mathbf{d}^c \oplus \tilde{\mathbf{e}}^c
\end{equation}
where $\mathbf{n^c} \in \mathbb{R}^{d_w+d_c}$. After obtaining the candidate news representation, it is utilized for user interest modeling and matching in the subsequent stages. 

\subsubsection{User Modeling}
The objective of utilizing this module was to derive user-interest representations from historical click records. We proposed a multi-granularity candidate-aware mechanism that interactively learned user interest representations at various levels of granularity from both the user’s clicked news and candidate news. Given that a user’s interests can often be diverse and only partially aligned with a candidate news item, this mechanism can address such variability~\citep{QiWW021,DBLP:conf/ijcai/LiuXWAX19}. Consequently, learning candidate news-aware user interest representations enhanced the modeling of user interests for better candidate news matching. Additionally, fine-grained interactions among different clicked news items could provide detailed clues for inferring user interest~\citep{FUM}. Different from previous studies that incorporated only news-level information, we fused word-, entity-, and news-level information for user interest modeling. 

\textbf{Word-level Granularity Candidate-aware Mechanism}. This section integrates the candidate news features into the word representation of a user’s clicked history news. Specifically, each word $t_{ijk}$ in the $j$-th genre news text of clicked news $c_i$ was mapped into a $d_w$-dimensional vector $\mathbf{t}_{ijk} \in \mathbb{R}^{d_w}$.
Different genres of news texts typically exhibit distinct semantic characteristics, which are crucial for a comprehensive understanding of news semantics~\citep{FUM}.  Therefore, the representation matrices of the user’s clicked news were stacked into an overall matrix $\mathbf{T} =[\mathbf{t}_{111}; \ldots; \mathbf{t}_{1gl}; \ldots; \mathbf{t}_{mgl}]$. For convenience, we rewrote $\mathbf{T}$ as follows:
\begin{equation}
    \mathbf{G} = [\mathbf{g}_1; \mathbf{g}_2; \dots; \mathbf{g}_L],~~ \mathbf{G} \in \mathbb{R}^{L\times d_w},~~L = mgl
\end{equation}
where $\mathbf{g}_i$ is the $i$-th word representation vector of the overall sequence of the clicked news. Although the Transformer network \citep{Transf} is widely recognized as a powerful method for document modeling, its efficiency in handling long documents is compromised by its quadratic complexity. Inspired by \citet{FUM}, Fastformer~\citep{Fastf} was utilized to model word-level interactions across extended sequences of user-clicked news. The fastformer’s core component, additive attention, enabled effective context modeling with linear complexity. In addition, to effectively learn the interaction between candidate news and clicked news at the word level, the query ($\mathbf{q}_i$), key ($\mathbf{k}_i$) and value ($\mathbf{v}_i$) were calculated as follows:
\begin{equation}\label{wc}
    \mathbf{q}_i = \mathbf{W}^w_q(\mathbf{g}_i \oplus \mathbf{n^c}),~~
    \mathbf{k}_i = \mathbf{W}^w_k\mathbf{g}_i,~~
    \mathbf{v}_i = \mathbf{W}^w_v\mathbf{g}_i
\end{equation}
where $\mathbf{W}^w_q \in \mathbb{R}^{d_w \times (2d_w+d_c)}$, $\mathbf{W}^w_k \in \mathbb{R}^{d_w \times d_w}$, and $\mathbf{W}^w_v \in \mathbb{R}^{d_w \times d_w}$ denote the trainable projection parameters. The candidate news-aware token representation $\hat{\mathbf{g}}_i$ can be calculated as follows:
\begin{equation}\label{wcan}
\begin{split}
    \mathbf{q} = Att(\mathbf{q}_1, \mathbf{q}_2, \dots, \mathbf{q}_L)\\
    \mathbf{k} = Att(\mathbf{q} * \mathbf{k}_1, \mathbf{q} * \mathbf{k}_2, \dots, \mathbf{q} * \mathbf{k}_L)\\
    \hat{\mathbf{g}}_i = \mathbf{W}_2(\mathbf{k} * \mathbf{v}_i)
\end{split}
\end{equation}
where ``$*$" denotes the element-wise product; $Att(\cdot)$ denotes the attention pooling network; and $\mathbf{W}_2$ represents the trainable parameters. Subsequently, the outputs of different attention heads were concatenated to construct a unified contextual representation $\bar{\mathbf{g}}_i$ for the $i$-th token. The word-level user interest representation $\mathbf{u}_w$ can be calculated as follows:
\begin{equation}
\mathbf{u} _w = softmax( \mathbf{W}_3 \cdot \mathbf{\bar{G}}^{T} + \mathbf{b}_3) \cdot \mathbf{\bar{G}}
\end{equation}
where $\mathbf{u} _w \in \mathbb{R}^{d_w}$, $\bar{\mathbf{G}}= [\bar{\mathbf{g}}_1; \bar{\mathbf{g}}_2; \ldots;  \bar{\mathbf{g}}_L] \in \mathbb{R}^{L \times {d_w}}$. $\mathbf{W}_3 \in \mathbb{R}^{d_w}$ and $\mathbf{b}_3 \in \mathbb{R}^{L}$ are the trainable parameters.

\textbf{Entity-level Granularity Candidate-aware Mechanism.} This mechanism can model user interest by incorporating candidate news into entity representations. We first obtained the initial entity representation $\mathbf{E}^u_{i} \in \mathbb{R}^{D \times d_e}$ for the \textit{i}-th clicked news item using TransE:
\begin{equation}
    \mathbf{E}^u_{i} = [\mathbf{e}^u_{i1}; \dots; \mathbf{e}^u_{ij}; \dots; \mathbf{e}^u_{iD}]
\end{equation}
where \textit{D} is the number of entities of each clicked news; and  $\mathbf{e}^u_{ij}$ is the representation of the \textit{j}-th entity of the \textit{i}-th clicked news. Similarly, entity representations for the entities in candidate news $\mathbf{E}^c = [\mathbf{e}^c_{1}; \mathbf{e}^c_{2}; \dots; \mathbf{e}^c_{D^c}] \in \mathbb{R}^{D^c \times d_e}$ were obtained. 
 To enhance the selection of informative relatedness between entities to match candidate news with user interests, a candidate-guided attention mechanism was designed to learn the match-aware representations for entities in news $c_i$. The relevance among entities in the \textit{i}-th clicked news $c_i$ and candidate news is first determined as follows:
\begin{equation}
    \mathbf{A}_i = \mathbf{E}^u_{i} \cdot (\mathbf{E}^c)^T 
\end{equation}
where $\mathbf{A}_i \in \mathbb{R}^{D \times D^c}$; and $\mathbf{A}_{ijk}$ denotes the relevance score between the \textit{j}-th entity of clicked news $c_i$ and the \textit{k}-th entity of the candidate news $n^c$. 
Then, the candidate news-weighted entity representation can be obtained as follows:
\begin{align}
    \tilde{\mathbf{e}}^u_{ij} = \sum_{k = 1}^{D^c} \beta_{jk}
    \mathbf{e}_{ij}^u\\
    \beta_{jk} = \frac{exp(\mathbf{A}_{ijk})}{\sum_{k=1}^{D^c} exp(\mathbf{A}_{ijk})}
\end{align}
where $\tilde{\mathbf{e}}^u_{ij} \in \mathbb{R}^{d_e}$; $exp(\cdot)$ is the exponential operation. Subsequently, the entity-level user-interest representation $\mathbf{u}_e \in \mathbb{R}^{d_e}$ is calculated as follows:
\begin{equation}\label{ec}
    \mathbf{u_e} =  MLP(\tilde{\mathbf{e}}^u_{11} \oplus \tilde{\mathbf{e}}^u_{12} \oplus \dots \oplus \tilde{\mathbf{e}}^u_{mD})
\end{equation}

\textbf{News-level Granularity Candidate-aware Mechanism.}  In this section, we first obtained the clicked news representations $\mathbf{d} = [\mathbf{d}_1, \mathbf{d}_2, \dots, \mathbf{d}_m]$ using the news content information extractor (see Section \ref{candid}), where $\mathbf{d}_i \in \mathbb{R}^{d_w}$ denotes the representation vector of the $i$-th user-clicked news. Subsequently, the transformer was used to capture interactive information from the user-clicked news. The multi-head self-attention mechanism, a key component of the transformer, was utilized. Unlike the standard transformer, we defined the query, key, and value for each head in the multi-head self-attention mechanism as follows:
\begin{equation}\label{nc}
\mathbf{q}_{i} = \mathbf{W}^n_q(\mathbf{d}_i \oplus \mathbf{n}^c),~ \mathbf{k}_{i} = \mathbf{W}^n_k\mathbf{d}_i,~
\mathbf{v}_{i} = \mathbf{W}^n_v\mathbf{d}_i
\end{equation}
where $\mathbf{W}^n_q \in \mathbb{R}^{d_w \times (2d_w+d_e)}$, $\mathbf{W}^n_k  \in \mathbb{R}^{d_w \times d_w}$, and $\mathbf{W}^n_v \in \mathbb{R}^{d_w \times d_w}$ are the learnable weight matrices of query, keys, and values, respectively. For each news item $c_i$, the self-attention learned a set of weights $\beta_i = \{\beta_{i1}, \beta_{i2}, \dots, \beta_{im}\}$ that that quantified the relevance of all clicked news \{$c_1$, $c_2$, $\dots$, $c_m$\} in relation to the query $\mathbf{q}_i$:
\begin{equation}
    \gamma_{ij} = \frac{exp(\mathbf{q}_i \mathbf{k}_j)}{\sum_{j'=1}^{m}exp(\mathbf{q}_i \mathbf{k}_{j'})}
\end{equation}
The output was obtained as a weighted sum of the values across all news items:
\begin{equation}
    \mathbf{f}_i = \sum_{j=1}^{m} \gamma_{ij}\mathbf{v}_j
\end{equation}
Subsequently, the outputs from the different attention heads were concatenated to form a unified contextual representation, which was then fed into a feed-forward network. We obtained each news representation $\mathbf{f}_i \in \mathbb{R}^{d_w}$, where $d_w$ denotes the vector dimension. The overall matrix of the news-level user-interest representation is denoted by $\mathbf{F}=[\mathbf{f}_1; \mathbf{f}_2; \cdots; \mathbf{f}_m]$, where $\mathbf{F} \in \mathbb{R}^{m \times d_w}$.
Finally, the news-level user interest representation can be calculated as follows:
\begin{equation}
    \mathbf{u}_n = softmax( \mathbf{W}_4 \cdot \mathbf{F}^{T} + \mathbf{b}_4) \cdot \mathbf{F}
\end{equation}
where $\mathbf{W}_4 \in \mathbb{R}^{d_w}$ and $\mathbf{b}_4 \in \mathbb{R}^m$ are trainable parameters.

As mentioned in the Introduction, interactions at different levels of granularity between candidate news and user-clicked news are crucial for user-interest modeling. Therefore, we combined word-, entity-, and news-level candidate-aware user interest representations to form the final user interest features:
\begin{equation}\label{ca}
    \mathbf{u} = \mathbf{u}_w \oplus \mathbf{u}_e \oplus \mathbf{u}_n
\end{equation}

\subsubsection{Candidate News with User Interests Matching}
Following previous work ~\citep{DBLP:conf/emnlp/WuWAQHHX19,DBLP:conf/ijcai/WuW0021},
the relevance $\hat{y} \in \mathbb{R}$ between user interests and candidate news was assessed using the dot product between the candidate news-aware user representation
$\mathbf{u}$ and the candidate news representation $\mathbf{n}^c$, represented as
$\hat{y} = (\mathbf{W}^u\mathbf{u})^T \cdot (\mathbf{W}^c\mathbf{n}^c)$, where $\mathbf{W}^u \in \mathbb{R}^{d \times (2d_w+d_e)}$ and $\mathbf{W}^c \in \mathbb{R}^{d \times (d_w+d_e)}$ are the trainable parameters. Candidate news was ranked based on their matching scores $\hat{y}$ for news recommendation.

Training dataset $D$ was generated using a negative sampling 
technique \citep{DBLP:conf/sigir/0001HDC18,DBLP:conf/ijcai/WuW0021}, where each positive sample was associated with $S$ negative samples randomly drawn from the same news impression. The loss function is defined using the BPR loss~\citep{DBLP:conf/uai/RendleFGS09}:

\begin{equation}\label{eatt}
\mathcal{L} =- \frac{1}{|D|}  {\textstyle \sum_{i=1}^{|D|}{\sigma (y_i^c-y_i^n)}}
\end{equation}
where $\sigma$ represents the sigmoid function; and 
$y_i^c$ and $y_i^n$ denote the matching scores for the $i$-th clicked and non-clicked news samples, respectively.

\section{Experiments and Results}\label{exper}
This section presents and analyzes the experimental results of the personalized news recommendation task, addressing the following \underline{r}esearch \underline{q}uestions.:
\begin{itemize}
    \item \textbf{RQ1}. How does the proposed MGCA model compare with other baseline methods for personalized news recommendations?
    \item \textbf{RQ2.} How does each module in MGCA affect the performance of personalized news recommendations?
    \item \textbf{RQ3.} What is the performance of MGCA with various hyperparameter settings?
    \item \textbf{RQ4.} Does MGCA provide improved news recommendations?
\end{itemize}

\subsection{Dataset and Experimental Settings}\label{subsection: ds}
The experiments were conducted using the MIND dataset \citep{DBLP:conf/acl/WuQCWQLLXGWZ20}, which has been extensively utilized in numerous studies on personalized news recommendations~\citep{TranSZTK23,FUM}. This dataset was constructed from six weeks of user behavior logs collected from Microsoft News between October 12 and November 22, 2019 included over twenty thousand impression logs. The logs from the final week were used for evaluation, whereas the remaining logs were used for model training and validation. Each impression log captured both the clicked and non-clicked news items shown to a user at a specific time, along with the prior news click behavior of the users. 
Following \citet{QiWW021}, entity embeddings were trained using the knowledge tuples extracted from WikiData using the TransE method~\citep{DBLP:conf/nips/BordesUGWY13}, and WikiData was employed as the knowledge graph in the experiments. 
Detailed statistics are listed in Table \ref{Data}.
The recommendation performance was evaluated using AUC, MRR, nDCG@5, and nDCG@10, following previous studies \citep{DBLP:conf/acl/AnWWZLX19,DBLP:conf/ijcai/WuWAHHX19}.

\begin{table}[!t]
\centering
\caption{Details of the MIND dataset.}
\label{Data}
\setlength{\tabcolsep}{2.5mm}{
\normalsize
\begin{tabular}{ll||ll}
\hline
Types & Number & Types & Number \\
\hline
\#Users & 94,057 & Average \#words in news title & 11.78 \\
\#News & 65,238 & Average \#entities in news title & 1.43 \\
\#Clicks & 347,727 & Average \#neighbors in KG & 18.21 \\
\#Impressions & 230,117 & Average \#clicks & 32.54 \\
\hline
\end{tabular}}
\end{table}

TensorFlow was used for model implementation. In our experiments, the word embeddings were 300-dimensional and initialized with GloVe embeddings~\citep{DBLP:conf/emnlp/PenningtonSM14}, considering only the first 30 words from news titles and the first five entities in the news articles. In addition, 10 neighbors for each entity were selected from the knowledge graph. Entity embeddings, represented as 100-dimensional vectors, were pretrained on the knowledge tuples extracted from WikiData using the TransE method~\citep{DBLP:conf/nips/BordesUGWY13}.
The numbers of heads in the candidate-aware Transformer and Fastformer mechanisms were set to 26 ($\lambda_1 = 26$) and 10 ($\lambda_2 = 10$), respectively\footnote{For convenience, let $\lambda_1$ and $\lambda_2$ represent the number of heads in candidate-aware Transformer and candidate-aware Fastformer mechanisms, respectively.}. The Adam optimizer was employed with a learning rate of 10-4, and training was conducted for 5 epochs with a batch size of 64 and a dropout probability of 0.2. All hyper-parameters were optimized using the validation dataset. 

\subsection{Comparison Baselines}\label{Comparison Baselines}
The proposed MGCA model was compared with the following baseline methods:
\begin{itemize}
  \item \textbf{DKN} \citep{DKNWangZXG18}: 
     This model dynamically aggregates news items clicked by a user and employs an attention mechanism to compute the attention weight between the candidate news and the user’s previously clicked news.

    \item \textbf{DAN} \citep{DBLP:conf/aaai/ZhuZSTG19}: This model employs a CNN to derive news representations from words and entities in news titles and utilizes an attentive LSTM network to capture user interest representations.

    \item \textbf{NAML} \citep{DBLP:conf/ijcai/WuWAHHX19}: It extracts news representations from titles, bodies, categories, and subcategories using multiple attentive CNNs.

    \item \textbf{NPA} \citep{DBLP:conf/kdd/WuWAHHX19}: Personalized attention networks are employed on words and previously clicked news to learn representations for both news and users.

    \item \textbf{LSTUR} \citep{DBLP:conf/acl/AnWWZLX19}: This model employs a GRU network to capture short-term user interests and uses user ID embeddings to represent long-term interests.

    \item \textbf{NRMS} \citep{WuWGQHX19}: This model utilizes multi-head self-attention mechanisms to derive representations for both news and users.

    \item \textbf{FIM} \citep{WangWLX20}: This matches user and candidate news by analyzing the text of users' clicked news and candidate news using a CNN.

    \item \textbf{KRED} \citep{DBLP:conf/recsys/LiuLWQCS020}: This model generates news representations by leveraging entities within the news and their associated neighbors in the knowledge graph through a graph attention network.
    
    \item  \textbf{FUM} \citep{FUM}: It models user interests from fine-grained behavioral interactions for news recommendations.

    \item \textbf{CupMar} \citep{TranSZTK23}: It leverages user profile representation across diverse contexts and multiple aspects of a news article to improve news recommendations.
\end{itemize}

\subsection{Main Results (RQ1)}\label{Results}

Each model was conducted five times, and the average performance of the various methods is presented in Table~\ref{mainresult}.  All values are shown as percentages, with the ``\%'' symbol omitted, and the best results are highlighted in bold. Table \ref{mainresult} presents the following observations. 

\begin{table}[!t]
\centering
\caption{Performance comparison of different models. The results are expressed in percentage.}
\label{mainresult}
\setlength{\tabcolsep}{6mm}{
\normalsize
\begin{tabular}{lcccc}
\hline
Models & AUC & MRR & nDGG@5 & nDGG@10 \\
\hline
\textbf{DKN} & 63.99 & 28.95 & 31.73 & 38.38 \\
\textbf{DAN} & 65.14 & 30.04 & 32.98 & 39.52 \\
\textbf{NAML} & 64.21 & 29.71 & 32.51 & 39.00 \\
\textbf{NPA} & 63.71 & 29.84 & 32.40 & 39.02 \\
\textbf{LSTUR} & 65.51 & 30.22 & 33.26 & 39.76 \\
\textbf{NRMS} & 65.36 & 30.02 & 33.11 & 39.61 \\
\textbf{FIM} & 64.46 & 29.52 & 32.26 & 39.08 \\
\textbf{KRED} & 65.61 & 30.63 & 33.80 & 40.23 \\
\textbf{FUM} & 66.81 & 31.37 & 34.81 & 41.22 \\
\textbf{CupMar} & 64.15 & 29.16 & 32.89  & 39.02 \\
\hline
\textbf{MGCA} & \textbf{67.07} & \textbf{31.71} & \textbf{35.16} & \textbf{41.50} \\
\hline
\end{tabular}}
\end{table}

The proposed MGCA model outperformed other baseline models such as FUM, FIM, and NRMS, which could not incorporate entity information or external knowledge. This superiority was attributed to the rich information provided by the entities and their neighbors in the knowledge graph, which could aid in inferring latent semantic details crucial for understanding user interests and candidate news. The results highlight the effectiveness of integrating knowledge graphs and entity information in modeling user preferences and news recommendations.

The proposed MGCA model demonstrated superior performance compared to baseline methods such as LSTUR, NRMS, and KRED, which modeled candidate news and user interests separately without considering their interrelationships. Independent modeling in these approaches complicated the accurate matching of user interests with candidate news. In contrast, our MGCA model introduced a multi-granularity, candidate-aware mechanism that interactively integrated user interests with candidate news. 

Third, MGCA outperformed the baseline models because the baseline methods were limited to capturing interactions at either the news or word level to model user interests. In contrast, MGCA utilized interactions at various levels of granularity between a user’s clicked news and candidate news, providing richer clues for user interest representation. Specifically, MGCA adopted a Transformer and Fastformer to capture relationship information at the news and word levels, respectively, while also incorporating external knowledge to capture rich entity information. Additionally, an attention mechanism was designed to capture the entity-level relationship between candidate news and user interests.

\subsection{Ablation Study (RQ2)}\label{Ablation}

To evaluate the impact of different components of our MGCA model, we conducted an ablation study by comparing several subnetworks. The ablation models used were as follows.
\begin{itemize}
\item \textbf{MGCA-w}: In this model, the word-level information was removed, meaning that the term $\mathbf{u}_e$ in Equation (\ref{ca}) was eliminated.
\item \textbf{MGCA-wc}: This removed the word-level granularity candidate mechanism, which means that $\mathbf{n^c}$ was eliminated in Equation (\ref{wc}).
\item  \textbf{MGCA-e}: IIn this model, the entity-level information was removed, meaning that $\mathbf{u}_e$ was omitted in Equation (\ref{ca}).
\item  \textbf{MGCA-ec}: It removed the entity-level granularity candidate awareness mechanism, using $\mathbf{e}_{ij}^u$ instead of $\tilde{\mathbf{e}}_{ij}^u$ in Equation (\ref{ec}).
\item  \textbf{MGCA-n}: This model removed the news-level information from user modeling, which means that we excluded $\mathbf{u}_n$ in Equation (\ref{ca}).
\item  \textbf{MGCA-nc}: This model removed the news-level granularity candidate mechanism from user modeling, which means that we excluded $\mathbf{n}^c$ in Equation (\ref{nc}).
\item \textbf{MGCA-c}: This model removed the word-, entity-, and news-level granularity candidate-aware mechanisms from user modeling.
\end{itemize}

\begin{table}[!t]
\centering
\caption{ Results of ablation experiment. The results are expressed in percentage.}
\label{resultecsp}
\setlength{\tabcolsep}{5.5mm}
{\normalsize
\begin{tabular}{lcccc}
\hline
Models & AUC & MRR & nDGG@5 & nDGG@10 \\
\hline
\textbf{MGCA-w}	& 66.44 & 31.27 & 34.66 & 41.04 \\
\textbf{MGCA-wc}& 66.53& 31.27& 34.70& 41.05 \\
\textbf{MGCA-e}	&66.08&31.21&34.63&40.98 \\
\textbf{MGCA-ec}	&66.61&31.59&34.97&41.24 \\
\textbf{MGCA-n}	& 65.96 &30.73 &34.19&40.54 \\
\textbf{MGCA-nc}& 66.16& 30.84&34.15&40.62 \\
\textbf{MGCA-c} & 65.55 & 30.88 & 34.11 & 40.50 \\
\hline
\textbf{MGCA} & 67.07 & 31.71 & 35.16 & 41.50\\
\hline
\end{tabular}
}
\end{table}

The ablation results are shown in Table~\ref{resultecsp}.
As expected, the performance of the ablated models consistently declined, clearly demonstrating the effectiveness of the proposed components. The findings were as follows. (1) Removing word-, entity-, or news-level information reduced the model’s performance across all four evaluation metrics. For instance, excluding news-level information resulted in decreases of 1.11\% in AUC, 0.98\% in MRR, 0.97\% in nDGG@5, and 0.96\% in nDGG@10. These results clearly demonstrated the importance of incorporating various granularities of information into news recommendation tasks. (2) Compared to MGCA-c, MGCA achieved a 1.52\% improvement in AUC, 0.83\% improvement in MRR, 1.05\% improvement in nDGG@5, and 1\% improvement in nDGG@10. This suggested that the multi-granularity candidate-aware mechanism significantly enhanced the news recommendation task. The effectiveness of this mechanism could arise from its ability to retain relevant information about candidate news, thereby minimizing the impact of irrelevant information on user modeling. (3) MGCA-wc, MGCA-ec, and MGCA-nc outperformed MGCA-c in most cases. Specifically, MGCA-wc exhibited improvements of 0.98\% in AUC, 0.39\% in MRR, 0.59\% in nDGG@5, and 0.55\% in nDGG@10 over MGCA-c. MGCA-ec achieved increases of 1.06\% in AUC and 0.71\% in MRR. MGCA-nc presented improvements of 0.04\% in nDGG@5 and 0.12\% in nDGG@10. These results highlight the importance of the three levels of candidate-aware mechanisms, demonstrating that each component of the MGCA model effectively captures crucial information regarding user interests.

\subsection{Inﬂuence of Hyper-parameters (RQ3)}
\begin{figure*}[!t]
\centering
\includegraphics[width= 0.8\columnwidth]{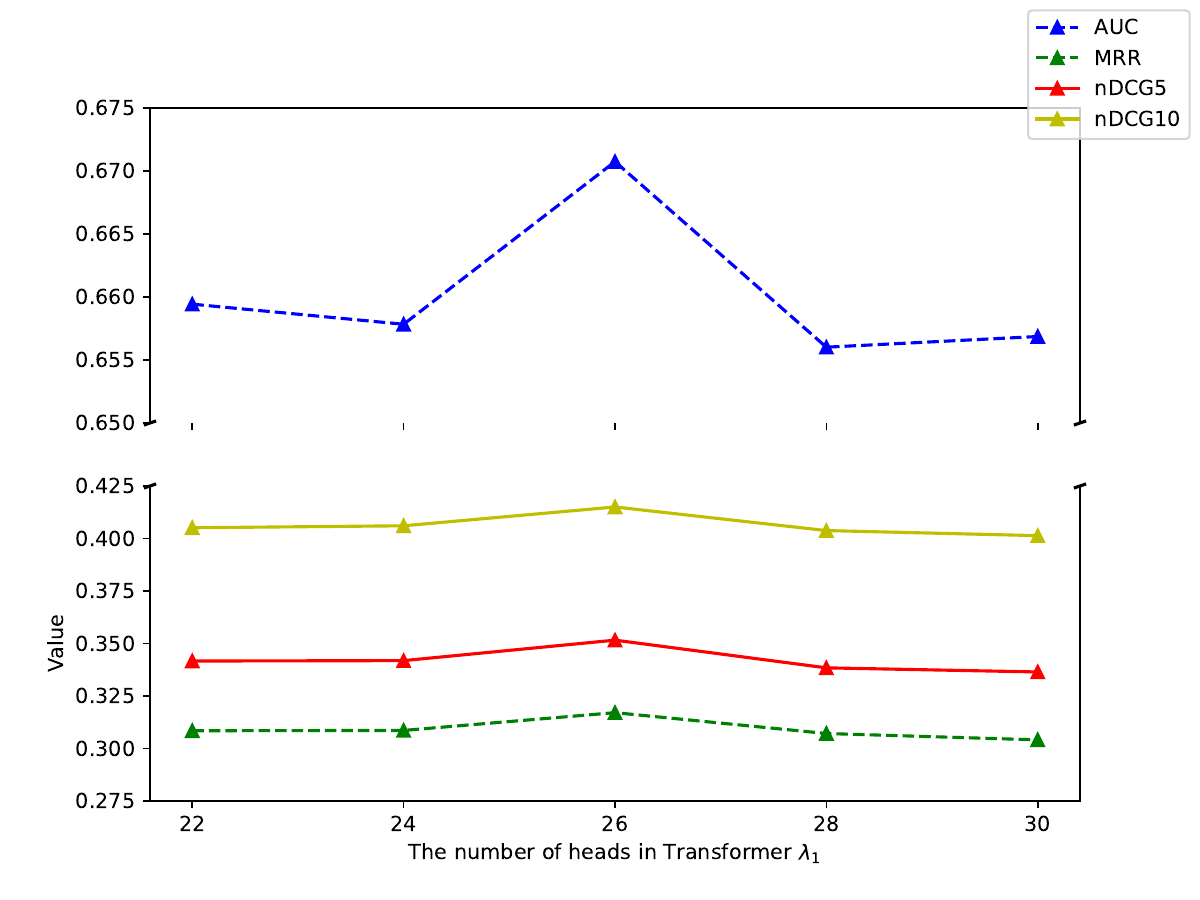} 
\caption{Performance of MGCA under diﬀerent number of self-attention heads in the Transformer.}
\label{figure4}
\end{figure*}

\begin{figure*}[!t]
\centering
\includegraphics[width= 0.8\columnwidth]{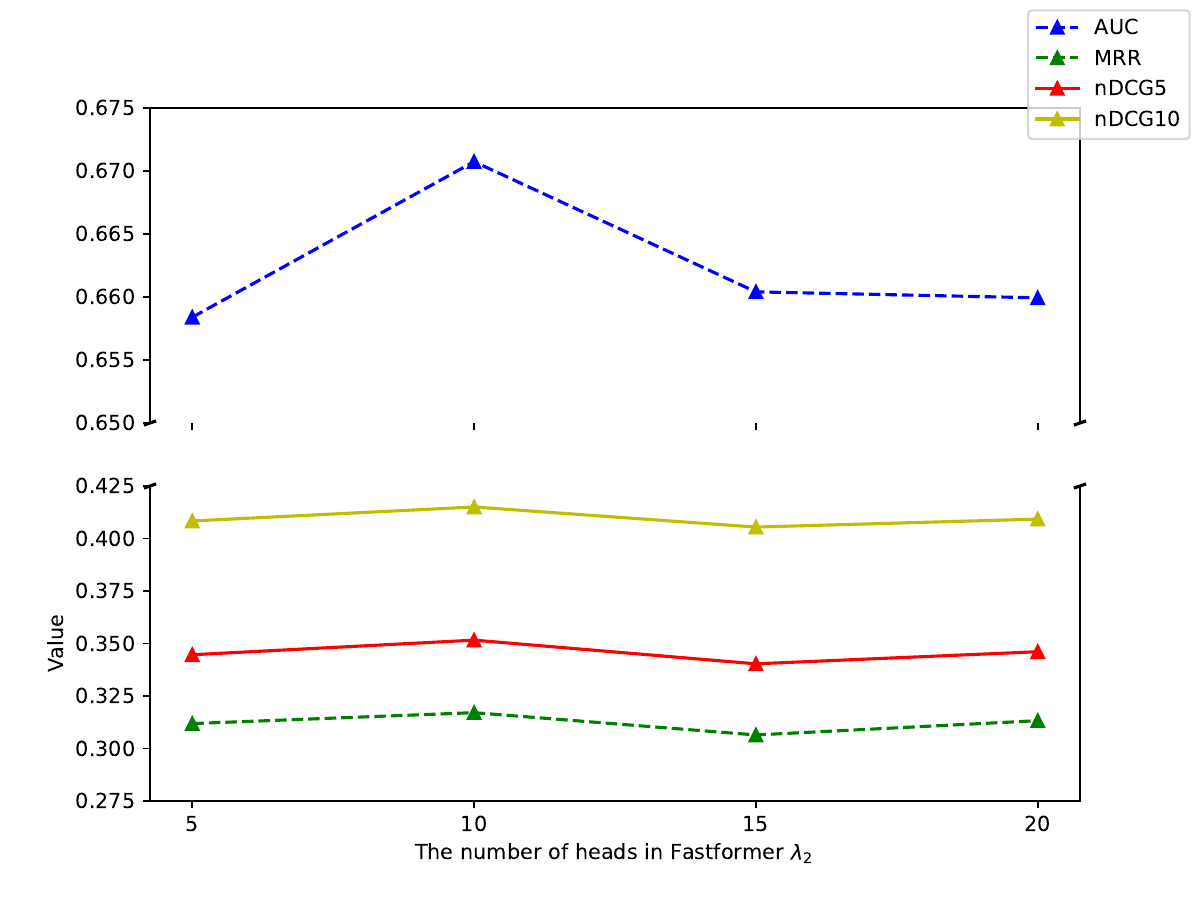} 
\caption{Performance of MGCA under diﬀerent number of self-attention heads in the Fastformer.}
\label{figure5}
\end{figure*}

In the Transformer and Fastformer modules, the number of self-attention heads could be a critical hyper-parameter for learning user interests (Figures~\ref{figure4} and ~\ref{figure5}). Two key observations could be achieved from these figures. First, the MGCA’s performance improved with an increasing number of self-attention heads because these heads helped capture valuable insights into the relationships between the entities in the clicked and candidate news, enhancing the understanding of user interests. Second, beyond a certain point, further increases in the number of self-attention heads led to performance degradation owing to the introduction of excessive parameters, which complicated the modeling of relationships between clicked and candidate news, potentially causing overfitting and reducing recommendation accuracy.

\subsection{Case Study (RQ4)}

A case study was conducted to assess the effectiveness of MGCA by comparing it with FUM, which was selected for comparison because of its superior performance relative to other baseline methods (Table \ref{mainresult}).


\begin{figure*}[!t]
\centering
\includegraphics[width= 1.02\columnwidth]{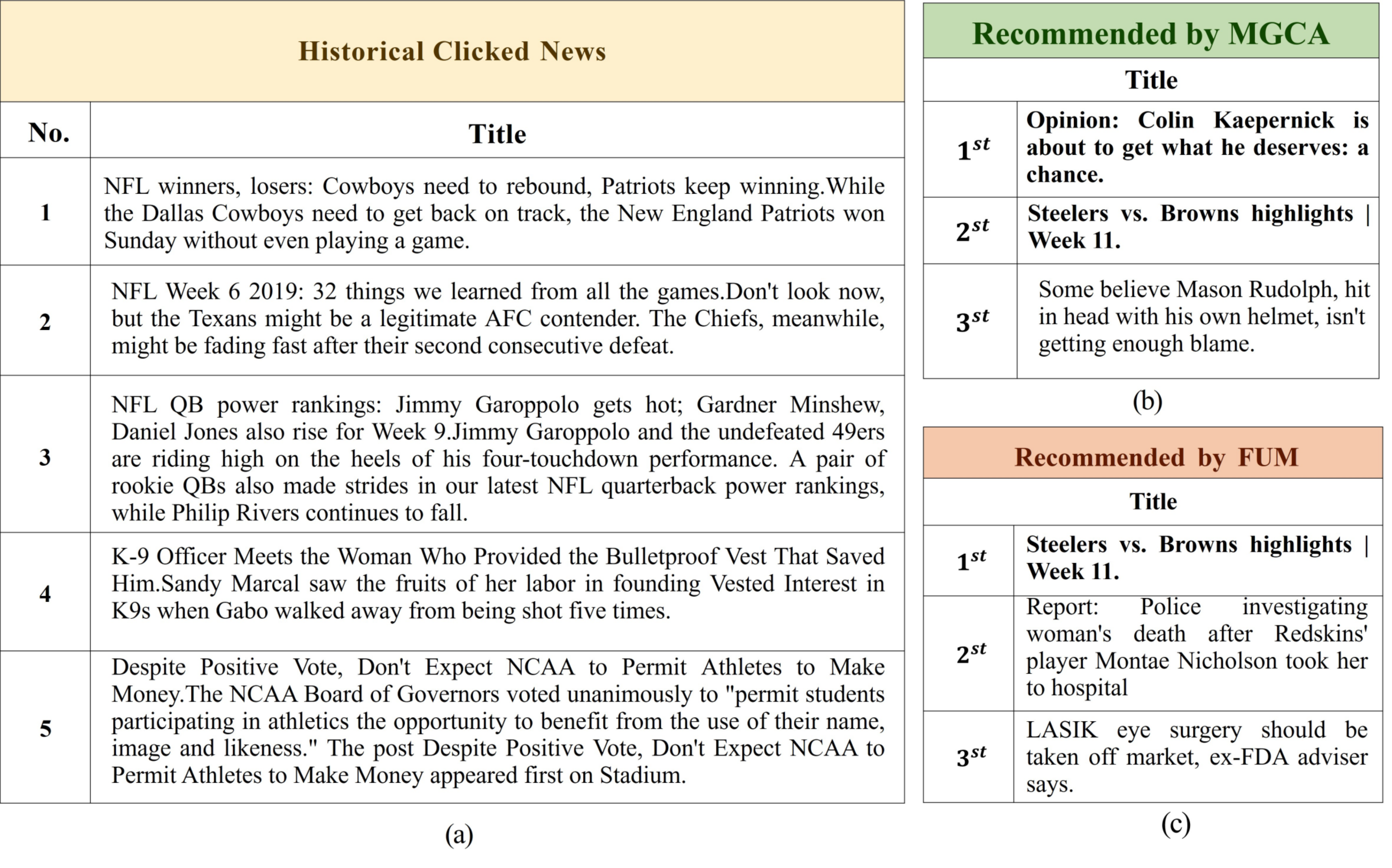} 
\caption{Case study on top 3 news recommended by MGCA and FUM in a sampled impression. The news actually clicked by the user is highlighted in bold.
}
\label{figure8}
\end{figure*}

Figure~\ref{figure8}(a) displays a sampled impression where the user had previously clicked on five news articles. The top three recommended news articles generated by MGCA and FUM are shown in Figures~\ref{figure8}(b) and~\ref{figure8}(c), respectively, with the articles clicked by the user highlighted in bold. From Figure~\ref{figure8}, the following conclusions can be drawn. (1) The user clicked on a news article about NFL (National Football League) games that was recommended by both the MGCA and FUM methods. This recommendation was based on the user’s prior engagement with two NFL-related articles (the first , second and third), indicating an interest in NFL game. Consequently, both MGCA and FUM utilized this information to recommend NFL-related news. (2) The user clicked on a news article titled  ``\textit{Opinion: Colin Kaepernick is about to get what he deserves: a chance}", which was recommended exclusively by MGCA. Since the first three news are related to ``\textit{NFL}" and Colin Kaepernick is a former NFL player, the MGCA model effectively identified the strong correlation between these articles and the recommended news using the knowledge graph and multi-granularity candidate-aware mechanism. In contrast, FUM that lacked the information specifically mentioning the ``\textit{Colin Kaepernick}'' struggled to accurately infer the user’s interest in this topic from the limited data available. This example further demonstrated the effectiveness of the proposed MGCA model.

\section{Conclusion}\label{conc}

This study introduced the \underline{m}ulti-\underline{g}ranularity \underline{c}andidate-\underline{a}ware (MGCA) framework for personalized news recommendations, which integrated information from multiple levels of granularity, including word-level, entity-level, and news-level, into user-interest modeling. The MGCA model employed a candidate-aware Transformer and Fastformer mechanisms, along with an attention mechanism, to capture the user interest representations at different granularities. These representations were then combined to create a comprehensive user interest profile, which was used to compute the relevance scores for candidate news articles. Extensive experiments conducted on a real-world dataset demonstrated that the MGCA model outperformed other baseline methods.

\section*{Acknowledgements}
This work is supported in part by the National Natural Science Foundation of China (Grant No. 62262045), the Humanity and Social Science Youth foundation of Ministry of Education of China (Grant No. 24YJCZH149), the Qingdao Postdoctoral Science Foundation (Grant No. QDBSH20230202038), the Open Research Foundation of Guangxi Key Lab of Human-machine Interaction and Intelligent Decision (Grant No. GXHIID2205).

\bibliography{ref}

\end{document}